# NOTE

# AN INTERPRETIVE FRAMEWORK FOR NARROWER IMMUNITY UNDER SECTION 230 OF THE COMMUNICATIONS DECENCY ACT

## INTRODUCTION

For well over a decade, courts and commentators have struggled to apply and interpret Section 230 of the Communications Decency Act of 1996 (CDA). Section 230 was designed to accomplish two objectives: First, Congress sought to protect children from Internet pornography by encouraging Internet Service Providers (ISPs) and websites to censor content voluntarily; second, Congress sought to promote freedom of expression on the Internet. To accomplish these two divergent goals, Section 230 grants immunity from tort liability to all websites and ISPs that are not themselves responsible for the creation or development of tortious content. Almost all courts have interpreted Section 230 immunity broadly, covering even publishers who take an active role in the production of controversial content, so long as they are not the authors. Although this broad interpretation effects the basic goals of the statute, it ignores several serious textual difficulties and mistakenly extends protection too far by immunizing even direct participants in tortious conduct. A proper reading of the statute—one that accounts for the background common law principles of vicarious tort liability upon which Section 230 was enacted—would correct both problems.

Part I introduces Section 230's history and purpose. Part II reviews the courts' broad interpretation of the statute. Part III examines several textual difficulties that this broad interpretation raises. Finally, Part IV attempts to solve these difficulties by interpreting Section 230 in light of the relationship between two strains of pre-Internet vicarious liability defamation doctrine and *Stratton Oakmont, Inc. v. Prodigy Services. Co.*, the defamation case that prompted Congress to pass Section 230. The analysis indicates that although the immunity provision of Section 230 is broad, Congress did not intend to abrogate traditional common law notions of vicarious liability. Some bases of vicarious liability remain, and their continuing validity both explains the interpretive difficulties and undergirds courts' recent push to narrow Section 230 immunity.

## I. SECTION 230: TEXT AND BACKGROUND

Congress enacted Section 230 of the CDA[1] to achieve two objectives:[2] to address the problem of children accessing pornography and other offensive material on the Internet,[3] and to promote freedom of expression on the Internet, a then-new and potentially fragile communications medium.[4] To accomplish these goals, Section 230 grants immunity from tort liability to computer service providers such as websites and ISPs that provide access to defamatory content created by third parties:

> § 230(c) Protection for "Good Samaritan" blocking and screening of offensive material
>
> > (1) Treatment of publisher or speaker
> >
> > No provider or user of an interactive computer service shall be treated as the publisher or speaker of any information provided by another information content provider.
> >
> > (2) Civil liability
> >
> > No provider or user of an interactive computer service shall be held liable on account of—
> >
> > (A) any action voluntarily taken in good faith to restrict access to or availability of material that the provider or user considers to be obscene, lewd, lascivious, filthy, excessively violent, harassing, or otherwise objectionable, whether or not such material is constitutionally protected;
>
> . . . .
>
> § 230(f)(3) Information content provider
>
> > The term "information content provider" means any person or entity that is responsible, in whole or in part, for the creation or development of information provided through the Internet or any other interactive computer service.[5]

---

[1]. Pub. L. No. 104-104, 110 Stat. 133 (1996).

[2]. *See* Zeran v. AOL, Inc., 129 F.3d 327, 330–31 (4th Cir. 1997) (reviewing the congressional intent behind Section 230).

[3]. *See* 47 U.S.C. § 230(b) (2006) ("It is the policy of the United States . . . to remove disincentives for the development and utilization of blocking and filtering technologies . . . ."); *Zeran*, 129 F.3d at 331 ("Another important purpose of § 230 was to encourage service providers to self-regulate the dissemination of offensive material over their services."); 141 CONG. REC. 15,503 (1995) (statement of Sen. Exon) ("[T]he worst, most vile, most perverse pornography is only a few click-click-clicks away from any child on the Internet.").

[4]. *See* 47 U.S.C. § 230(a)(3) (2006) ("The Internet and other interactive computer services offer a forum for a true diversity of political discourse, unique opportunities for cultural development, and myriad avenues for intellectual activity."); *Zeran*, 129 F.3d at 330 ("Congress recognized the threat that tort-based lawsuits pose to freedom of speech in the new and burgeoning Internet medium.").

It is easy to see how Congress's grant of immunity promotes freedom of speech: Internet-based publishers are enabled to relay and distribute controversial and even defamatory third-party-created content without fear of tort liability. So long as they are not the authors of the material, "information content providers"—websites and other service providers—will not be liable. As a result, Comcast and Verizon can provide unfettered access to the entire Internet, blog-hosting sites such as WordPress and Blogger can make available the enlightening and sometimes less-enlightening musings of the Internet community at large, and Wikipedia can provide user-authored, encyclopedic coverage of nearly every topic imaginable—all free from the threat of liability should some user-submitted content prove to be defamatory.[6]

---

5. 47 U.S.C. § 230(c), (f)(3) (2006).

Section 230's effect on children's access to objectionable content is slightly more roundabout. Rather than creating a positive incentive to censor content, the provision operates by removing a major disincentive to censorship: the threat of defamation liability. In *Stratton Oakmont, Inc. v. Prodigy Services Co.*,[7] a New York state court held Prodigy, a then-popular ISP, liable for defamatory content posted by a third party to one of the service's message boards. The court reasoned that because Prodigy held itself out to the public as a family-friendly, carefully controlled and edited Internet provider, and took steps to screen offensive content, the ISP had taken on the role of a newspaper-like publisher rather than a mere distributor[8] and could therefore be held liable.[9] By filtering some objectionable content, the court reasoned, Prodigy had taken ownership of all of it. Congress rejected this line of reasoning in Section 230 and instead immunized computer service providers from liability "on account of any action voluntarily taken . . . to restrict access to or availability of [objectionable content]."[10] As a result, censorship of third-party-created content can now proceed freely. Without fear of incurring liability, filtered Internet services can protect homes and workplaces from pornography and other objectionable material, message board administrators can remove obscene or simply off-topic posts from their sites, and bloggers can remove or censor objectionable visitor comments to their postings.

---

6. As mere hosts of content created by others, blog-hosting sites are immune from defamation liability under a straightforward application of Section 230. For a thoughtful discussion of the slightly more difficult question of the applicability of Section 230 immunity to Wikipedia, see Ken S. Myers, *Wikimmunity: Fitting the Communications Decency Act to Wikipedia*, 20 HARV. J.L. & TECH. 163 (2006).

7. No. 31063/94, 1995 WL 323710 (N.Y. Sup. Ct. May 24, 1995).

8. Distributors, such as libraries, bookstores, and telephone companies, who "deliver, transmit, or facilitate defamation [but] have only the most attenuated or mechanical connection with the defamatory content" are not liable "unless [they] know[] or should know of the defamatory content." Publishers and republishers, such as book presses and newspapers, on the other hand, are responsible for all harms caused by their defamatory publications. DAN B. DOBBS, THE LAW OF TORTS § 402 (2000).

9. 1995 WL 323710, at *4.

10. *See* 47 U.S.C. § 230(c)(2) (2006).

## II.　Z*ERAN*, D*RUDGE*, AND THE MAJORITY VIEW

Section 230 operates straightforwardly: Computer service providers such as ISPs and websites are granted immunity from defamation liability for third-party-created content in order to promote free speech and to allow them to protect children from objectionable content. Courts have struggled, however, to define the precise contours of the statute's immunity provisions. What qualifies as an interactive computer service? Can a business or website simultaneously be both a computer service and a content provider? Can a website or other service provider ever edit content so heavily as to transform itself into a content provider and thereby lose its immunity?

Courts have, from the beginning, adopted a broad view of Section 230 immunity.[11] In *Zeran v. America Online, Inc.*,[12] the U.S. Court of Appeals for the Fourth Circuit, the first circuit to interpret the statute, held that even if a service provider exercised significant editorial control over the content in question, it was immune so long as it was not the content's author.[13] Later courts gradually expanded Section 230 immunity to cover an increasingly broad range of potential defendants.[14] In *Blumenthal v. Drudge*,[15] for ex-

---

11. *See* Mark A. Lemley, *Rationalizing Internet Safe Harbors*, 6 J. TELECOMM. & HIGH TECH. L. 101, 103 (2007) ("[Section 230] has been interpreted quite broadly to apply to any form of Internet intermediary, including employers or other companies who are not in the business of providing Internet access and even to individuals who post the content of another. And it has been uniformly held to create absolute immunity from liability for anyone who is not the author of the disputed content, even after they are made aware of the illegality of the posted material and even if they fail or refuse to remove it." (footnotes omitted)).

12. 129 F.3d 327 (4th Cir. 1997).

13. *See id.* at 330 ("[L]awsuits seeking to hold a service provider liable for its exercise of a publisher's traditional editorial functions—such as deciding whether to publish, withdraw, postpone or alter content—are barred."). The plaintiff in *Zeran* alleged that America Online had unreasonably delayed in retracting defamatory messages posted by a third party, refused to post retractions to those messages, and failed to screen for and prevent similar future postings. *Id.* at 328. The court rejected these arguments, interpreting the text of Section 230 to preclude liability even where a service provider is on notice of and in a position to prevent or remove potentially defamatory content. The entire inquiry turned on the identity of the content's author. *Id.* at 330–32 ("By its plain language, § 230 creates a federal immunity to any cause of action that would make service providers liable for information originating with a third-party user of the service.").

14. For a useful table of cases illustrating the expanding scope of Section 230 immunity, see Myers, *supra* note 6, at 205–08. *See also* H. Brian Holland, *In Defense of Online Intermediary Immunity: Facilitating Communities of Modified Exceptionalism*, ample, the U.S. District Court for the District of Columbia found an ISP eligible for Section 230 immunity even though it had contracted for the development of the unverified gossip column that was at issue in the suit.[16] Thus, even when defamatory content is developed at a service provider's request, the provider is immune from liability so long as it is not the author of the material.

Almost all courts considering Section 230's scope have followed *Zeran* and *Drudge*,[17] and the Ninth Circuit has formalized the holdings of the cases into a three-part inquiry:[18] (1) is the defendant an "interactive computer service" within the meaning of Section 230; (2) does the plaintiff's cause of action treat the defendant as a publisher; and (3) was the content at issue in the suit "provided by another information content provider?" If a plaintiff's cause of action against a website or other computer service treats that service as a publisher of third-party-created content, the defendant will be immune from liability—end of story.

---

56 U. KAN. L. REV. 369, 374 (2008) ("Following *Zeran*, and building on that court's reading of both the statute and the policies sought to be effected, courts have consistently extended the reach of § 230 immunity along three lines: (1) by expanding the class who may claim its protections; (2) by limiting the class statutorily excluded from its protections; and (3) by expanding the causes of action from which immunity is provided."); Brandy Jennifer Glad, Comment, *Determining What Constitutes Creation or Development of Content Under the Communications Decency Act*, 34 SW. U. L. REV. 247, 253–58 (2004) (individually discussing several cases in the "series of decisions that offered increasingly broader immunity for ISPs").

15. 992 F. Supp. 44 (D.D.C. 1998).

16. *Id.* at 47, 50.

17. *See* Universal Commc'n Sys., Inc. v. Lycos, Inc., 478 F.3d 413, 418 (1st Cir. 2007); Batzel v. Smith, 333 F.3d 1018, 1027 (9th Cir. 2003); Green v. AOL, Inc., 318 F.3d 465, 471 (3d Cir. 2003); Ben Ezra, Weinstein, & Co. v. AOL, Inc., 206 F.3d 980, 986 (10th Cir. 2000). *But see* Chi. Lawyers' Comm. for Civil Rights Under Law, Inc. v. Craigslist, Inc., 519 F.3d 666, 670 (7th Cir. 2008). Judge Easterbrook reasons in *Craigslist* that other circuits have extended immunity too far by "treating § 230(c)(1) as a grant of comprehensive immunity from civil liability for content provided by a third party" and urges instead immunity only for publication-based torts. *Id.* Because publication is not an element of copyright infringement, Judge Easterbrook's interpretation would allow, for example, computer service providers to be held liable for contributory copyright infringement if their systems were designed to help people steal music.

18. *See Batzel*, 333 F.3d at 1037 (Gould, J., concurring in part and dissenting in part).

### III. INTERPRETIVE DIFFICULTIES AND OUTMODED POLICY OBJECTIVES

The prevailing approach to Section 230 serves Congress's dual goals quite well. The broad service provider immunity of *Zeran* and *Drudge* promotes freedom of speech on the Internet while simultaneously removing a major disincentive to censorship. *Zeran* and *Drudge*, along with later cases, however, failed to resolve several serious interpretive difficulties.

#### A. Subsections 230(c)(1) and (c)(2): Giving Meaning to "Good Samaritan"

Chief among these difficulties is the relationship between subsections (c)(1) and (c)(2). Subsection (c)(1) provides that "[n]o provider or user of an interactive computer service shall be treated as the publisher or speaker of any information provided by another information content provider."[19] As noted in Part II above, *Zeran* and later courts quite reasonably interpreted this language to immunize websites and other service providers from tort liability for any third-party-created content regardless of whether they make any editorial changes to the content.[20] If this reading is proper, however, then subsection (c)(2), which provides immunity to computer service providers that do choose to censor objectionable third-party-created content, appears to be superfluous.[21] If providers who choose to censor third-party-created content are already immune under subsection (c)(1) because the content is not their own, then what can be the purpose of subsection (c)(2), which grants immunity if they choose to censor? The rule against surplusage,[22] which would apply with particular force here where the two subsections are textual neighbors, thus militates against the majority view of *Zeran* and *Drudge*. Similarly, the very heading of Section 230 and the title of the act of which it is a part coun-sel against the majority view: "§ 230(c)—which is, recall, part of the 'Communications Decency Act'—bears the title 'Protection for "Good Samaritan" blocking and screening of offensive material', hardly an apt description if its principal effect is to induce ISPs to do nothing about the distribution of indecent and offensive material."[23] The majority view reads subsection (c)(2) entirely out of the text and in the process renders the Section powerless to achieve its stated objective—encouragement of self-censorship.[24]

---

19. 47 U.S.C. § 230(c)(1) (2006).

20. *See Batzel*, 333 F.3d at 1037.

21. Subsection (c)(2) reads: "No provider or user of an interactive computer service shall be held liable on account of—(A) any action voluntarily taken in good faith to restrict access to [objectionable content]." 47 U.S.C. § 230(c)(2) (2006).

22. The rule against surplusage presumes that Congress does not include redundant or otherwise unnecessary language in statutes. WILLIAM N. ESKRIDGE, JR. ET AL., LEGISLATION AND STATUTORY INTERPRETATION 266 (2000); *see also* Kungys v. United States, 485 U.S. 759, 778 (1988) (calling it the "cardinal rule of statutory interpretation that no provision should be construed to be entirely redundant").

### B. Subsections 230(c)(1) and (f)(3): Defining Information Content Providers

The language of Section 230 implies a world composed of two distinct categories of entities: information content providers and computer service providers. Information content providers are liable in tort for the damages their content causes, whereas computer service providers such as ISPs and websites are immune. But what happens when the distinction between service provider and content provider becomes blurred? Is it possible for multiple individuals to be collectively responsible for content? Could an ISP or website be simultaneously both a service provider and a content provider? Without clear statutory guidance, courts[25] and commentators[26] have struggled mightily to discover

---

23. Doe v. GTE Corp., 347 F.3d 655, 660 (7th Cir. 2003) (Easterbrook, J.).

24. *See id.* (noting that if the majority approach is correct, "then *§ 230(c)* as a whole makes ISPs indifferent to the content of information they host or transmit: whether they do (*subsection (c)(2)*) or do not (*subsection (c)(1)*) take precautions, there is no liability under either state or federal law" (emphasis added)).

25. *See, e.g.*, Fair Hous. Council of San Fernando Valley v. Roommates.com, LLC, 521 F.3d 1157, 1184 (9th Cir. 2008) (searching dictionaries, case law, and the text of the statute for the "ordinary meaning" of "development"); *Batzel*, 333 F.3d at 1031 ("The 'development of information' therefore means something more substantial than merely editing portions of an e-mail and selecting material for publication."); Ben Ezra, Weinstein & Co. v. AOL, Inc., 206 F.3d 980, 984–86 (10th Cir. 2000) (holding defendant not liable when plaintiff alleged defendant was acting as both computer service and content provider); Anthony v. Yahoo!, Inc., 421 F. Supp. 2d 1257, 1262–63 (N.D. Cal. 2006) (stating that even when Yahoo! did not create the online material, the CDA "[did] not absolve Yahoo! from liability for any accompanying misrepresentations" they made that the material was genuine); MCW, Inc. v. Badbusinessbureau.com, L.L.C., No. Civ. A.3:02-CV-2727-G, 2004 WL 833595, at *8 (N.D. Tex. Apr. 19, 2004) ("The distinction between merely publishing information provided by a third-party as an interactive computer service and actually creating or developing any of the information posted as an information content provider is critical [and] determines whether the CDA provides immunity."); Carafano v. Metrosplash.com, Inc., 207 F. Supp. 2d 1055, 1066–68 (C.D. Cal. 2002) (finding that defendant, although a computer service, was also an in-the precise distinction between a mere service provider that exercises traditional editorial discretion and an information content provider, receiving no immunity under Section 230.[27]

Consider the facts before the court in *MCW, Inc. v. Badbusinessbureau.com, L.L.C.*[28] In *MCW* the U.S. District Court for the Northern District of Texas considered a claim against Badbusinessbureau.com (BBB),[29] a web-based forum that allows consumers to post business complaints and vent their frustrations to the Internet-browsing public.[30] MCW, the target of several negative user-submitted reviews, brought an action against BBB, alleging a variety of common law and statutory violations.[31] BBB sought dismissal, claiming immunity under Section 230.[32] If BBB were a mere conduit through which customers could publicize their grievances, the case would have been simple. Such conduits clearly are immune as service providers un-

---

formation content provider, and thus could not invoke Section 230 immunity); Blumenthal v. Drudge, 992 F. Supp. 44, 50–51 (D.D.C. 1998) (discussing the effects of a licensing agreement between AOL and Drudge on AOL's immunity under Section 230); Gentry v. eBay, Inc., 121 Cal. Rptr. 2d 703, 714–18 (Cal. Ct. App. 2002) (concluding that plaintiff failed to prove that eBay acted "outside the immunity for service providers").

26. *See, e.g.*, Robert G. Magee & Tae Hee Lee, *Information Conduits or Content Developers? Determining Whether News Portals Should Enjoy Blanket Immunity from Defamation Suits*, 12 COMM. L. & POL'Y 369 (2007); Bryan J. Davis, Comment, *Untangling the "Publisher" Versus "Information Content Provider" Paradox of 47 U.S.C. § 230: Toward a Rational Application of the Communications Decency Act in Defamation Suits Against Internet Service Providers*, 32 N.M. L. REV. 75 (2002); Glad, Comment, *supra* note 14.

27. *Zeran* and later decisions have made it quite clear that a computer service provider need not be a mere conduit in order to qualify for immunity. So long as its actions do not take it out of the traditional realm of editorial discretion and into the realm of authorship, a computer service will retain its Section 230 immunity. Zeran v. AOL, Inc., 129 F.3d 327, 330 (4th Cir. 1997) ("[Section] 230 precludes courts from entertaining claims that would place a computer service provider in a publisher's role. Thus, lawsuits seeking to hold a service provider liable for its exercise of a publisher's traditional editorial functions—such as deciding whether to publish, withdraw, postpone or alter content—are barred."). Of course, at the extreme, the exercise of traditional editorial functions can start to look very much like co-authorship.

28. No. Civ. A.3:02-CV-2727-G, 2004 WL 833595 (N.D. Tex. Apr. 19, 2004).

29. The defendant's website, "The Ripoff Report," can be accessed at both http://www.badbusinessbureau.com and http://www.ripoffreport.com.

30. *MCW*, 2004 WL 833595, at *1.

31. The plaintiff alleged, among other violations, unfair competition, business disparagement, and trademark infringement. *Id.* at *2.

32. *Id.* at *7.

der Section 230. But the site was much more than a conduit. After receiving complaints, the site operators categorized them by geographic region and often added disparaging titles.[33] On some occasions the site had actively solicited the negative content submissions from its users,[34] further blurring the line between service provider and content provider.

The court was forced to decide whether BBB was merely a service provider that had chosen to exercise a rather broad degree of editorial discretion or whether it had crossed the provider-creator line and become responsible for the content in question.[35] This line is difficult to draw. Unfortunately, the text of Section 230 provides little guidance. Section 230(f)(3) defines an information content provider as "any person or entity that is responsible, in whole or in part, for the creation or development of information provided through the Internet or any other interactive computer service."[36] Does soliciting content make one "responsible for its creation"? Does adding a title to another's content make one a partial creator of that content? Perhaps, but perhaps not.[37]

The *Zeran* line of cases does little to clarify the standard, for they simply restate the statutory language while ignoring the potential for ambiguity at the margins.[38] When confronted with facts that force a resolution of the ambiguous distinction between service provider and content provider, courts almost un-failingly resolve the issue in favor of immunity.[39] Unless a service provider literally and unambiguously pens the words of the content in question, it will be immune from liability.

---

33. *Id.* at *1, *9 nn.10–11.

34. *Id.* at *10.

35. *Id.* at *7–8 ("The CDA requires courts to determine . . . when content provided by third-parties is somehow transformed into content created or developed by an interactive computer service. The distinction between merely publishing information provided by a third-party as an interactive computer service and actually creating or developing any of the information posted as an information content provider is critical." (citation omitted)).

36. 47 U.S.C. § 230(f)(3) (2006).

37. The *MCW* court ultimately concluded that BBB was not entitled to protection under Section 230. *Id.* at *10. Its conclusion was based in part on the argument that even if BBB was not literally the creator or developer of the content at issue, it was at the very least "*responsible* . . . for the creation or development" of that content. *Id.* at *10 n.12. This interpretation of Section 230(f)(3)'s definition of information content provider is broader than the majority view, narrowing Section 230 immunity. It provides, however, no more guidance than does the majority view for determining where to draw the line between mere service providers and content providers. It simply moves that line in a direction less favorable to immunity.

38. Zeran v. AOL, Inc., 129 F.3d 327, 330 (4th Cir. 1997) ("[Section] 230 creates a federal immunity to any cause of action that would make service providers liable for information originating with a third-party user of the service.").

Followed to its logical conclusion, this view would immunize parties surely not within the intended scope of Section 230. Imagine, for example, a hypothetical website, harassthem.com.[40] Visitors to the site are encouraged to get even with others by publicly posting a target's name, address, credit card information, and so forth, along with embarrassing facts or stories about him.[41] The site instructs users that the information need not be confirmed and can be based on rumor, conjecture, or fabrication. Such a site, by providing a forum for and encouraging defamation, can be quite reasonably considered "responsible . . . in part, for the creation or development"[42] of the resulting tortious content, rendering it ineligible for Section 230 immunity. Yet the content is not authored by harassthem.com. It "originat[es] with a third-party user of the service,"[43] which, under *Zeran*, is all that is necessary for service provider immunity. *Zeran* fails to account for those circumstances in which immunity should be cut off even absent full authorship. Consequently, its framework is of little use in drawing the appropriate line between mere service providers and those responsible for tortious content.

### C.  Now-Undesirable Policy Outcomes

The *Zeran* line also suffers from a flaw not of its own creation. Section 230 takes as an explicit objective the "continued development of the Internet . . . [and preservation of] the vibrant and competitive free market that presently exists for the Internet."[44]

---

39. *See, e.g.*, Carafano v. Metrosplash, Inc., 339 F.3d 1119, 1120–21 (9th Cir. 2003); Green v. AOL, Inc., 318 F.3d 465, 468 (3d Cir. 2003); Ben Ezra, Weinstein, & Co. v. AOL, Inc., 206 F.3d 980, 983 (10th Cir. 2000); Novak v. Overture Servs., Inc., 309 F. Supp. 2d 446, 452–53 (E.D.N.Y. 2004); Blumenthal v. Drudge, 992 F. Supp. 44, 50 (D.D.C. 1998); Doe v. AOL, Inc., 783 So. 2d 1010, 1018 (Fla. 2001).

40. This example is from Fair Housing Council of San Fernando Valley v. Roommates.com, LLC. 489 F.3d 921, 928 (9th Cir. 2007).

41. The scenario is, unfortunately, not all that far-fetched. Consider sites such as DontDateHimGirl.com, ManHaters.com, and TrueDater.com that offer disgruntled lovers the opportunity to vent their frustrations and warn future victims of their ex-partners' faults, mixing public service and sweet revenge. *See* Lizette Alvarez, *(Name Here) Is a Liar And a Cheat*, N.Y. TIMES, Feb. 16, 2006, at G1.

42. 47 U.S.C. § 230(f)(3).

43. *Zeran*, 129 F.3d at 330.

44. 47 U.S.C. § 230(b) (2006).

Congress recognized the enormous potential of the Internet and feared that unchecked tort liability might decrease its value as a facilitator of free speech.[45] But, as some have noted, "the Internet is no longer in its infancy."[46] Internet publications have matured to the point where, at least in certain instances, they are robust enough to face the same exposure to liability as their print counterparts.[47] For example, it no longer serves any coherent purpose to treat defamatory content in the print edition of the *New York Times* differently than that in the online version.[48] This is not to say, of course, that the Internet has no need for protection. Certain contexts may very well warrant special protections. Many continue to defend vigorously the special protection that Section 230 currently affords Internet publications because of the Internet's unique status as a facilitator of free individual expression.[49] Still, the Internet landscape has changed dramatically since *Zeran* was decided, and its extremely broad immunity seems slightly out of step with modern policy objectives. The Internet continues to serve as a valuable facilitator of free expression, but it has now become such a robust and integral part of modern life that it can safely be subjected to at least minimal regulation. Indeed, the ubiquity of modern Internet-based communication suggests that such regulation is not only feasible but normatively desirable. As an increasing percentage of human interaction and communication is transferred to the digital realm, legal remedies must follow, lest familiar wrongs be left without familiar remedies.[50]

---

45. *Zeran*, 129 F.3d at 330.

46. Jae Hong Lee, Note, *Batzel v. Smith & Barrett v. Rosenthal: Defamation Liability for Third-party Content on the Internet*, 19 BERKELEY TECH. L.J. 469, 491 (2004) ("[T]he Internet is no longer in its infancy, having grown into a vigorous and muscular adolescent.").

47. *See* Magee & Lee, *supra* note 26 (arguing that the same defamation standard should apply to both print and web-based news portals).

48. Newspapers author most of their material and so are not in a position to benefit from Section 230 immunity. Third-party-authored advertisements, though, are the exception. In the print context, courts have held primary publishers like newspapers responsible even for advertisements prepared by others. *See* DOBBS, *supra* note 8, § 402 (citing Triangle Pubs., Inc. v. Chumley, 317 S.E.2d 534 (Ga. 1984)). Online newspapers, however, would be immune from liability under Section 230.

49. *See, e.g.*, Holland, *supra* note 14, at 391–404.

50. *See Hearing on Cyberbullying and other Online Safety Issues for Children; H.R. 1966, the "Megan Meier Cyberbullying Prevention Act"; and H.R. 3630, the "Adolescent Web Awareness Requires Education Act (AWARE Act) Before the H. Comm. on the Judiciary*, 111th Cong. (2009), *available at* http://judiciary.house.gov/hearings/hear_090930.html (testimony of John Palfrey, Harvard Law Sch.) ("[N]ewly criminalizing a broad

IV. A NARROWER VIEW OF SECTION 230 IMMUNITY

The search for coherence and the desire to achieve sensible policy outcomes have led a growing number of courts to apply Section 230 immunity more narrowly. By interpreting "content development" to encompass more than mere literal authorship, some courts have allowed plaintiffs' claims to go forward against websites even where a third party created the content at issue.

For example, in *Fair Housing Council of San Fernando Valley v. Roommates.com, LLC*,[51] the Ninth Circuit considered a discrimination claim against the website Roommates.com. The site was designed to match people renting spare rooms with people looking for housing.[52] A user posting an available housing opportunity on the site was required to provide not only basic identificatory information but also more personal and controversial information such as his sex, sexual orientation, and whether he would bring children to the household.[53] The user was asked to provide this data by selecting from a limited set of answer choices.[54] The Fair Housing Councils of San Fernando Valley and San Diego

---

swath of online speech is not the right general approach. Nor do I favor a set of rules that apply only in cyberspace and not in offline life. The rules should, to the greatest extent possible, be the same in the online context as offline. We should strive to apply rules of general applicability to the Internet context."); JOHN PALFREY & URS GASSER, BORN DIGITAL: UNDERSTANDING THE FIRST GENERATION OF DIGITAL NATIVES 106–07 (2008) ("The scope of the immunity the CDA provides for online service providers is too broad. . . . There is no reason why a social network should be protected from liability related to the safety of young people simply because its business operates online."); *see also* Susan Freiwald, *Comparative Institutional Analysis in Cyberspace: The Case of Intermediary Liability for Defamation*, 14 HARV. J.L. & TECH. 569, 654 (2001) ("[T]otal immunity for intermediaries rather than distributor liability represents a failure of public policy and the poor resolution of a legal conflict."); Lemley, *supra* note 11, at 101–02 (arguing that Section 230's immunity provision is inconsistent with and less desirable than other federal Internet intermediary safe harbors); Melissa A. Troiano, Comment, *The New Journalism? Why Traditional Defamation Laws Should Apply to Internet Blogs*, 55 AM. U. L. REV. 1447, 1475 (2006) ("Because many bloggers utilize their blogs to attract a large public audience in a way that resembles the function of traditional print journalism, bloggers should not be immune from suit simply because they publish their work on the Internet. Instead, bloggers who choose to share their views with the public, and who individually monitor their content, should be responsible for ensuring the legality of their content prior to publication.").

51. 521 F.3d 1157 (9th Cir. 2008).
52. *Id.* at 1161.
53. *Id.* at 1161–62.
54. *Id.* at 1166.

sued, alleging that the site violated the federal Fair Housing Act and California housing discrimination laws.[55]

Writing for a divided Ninth Circuit panel, Chief Judge Kozinski rejected the defendant's Section 230 immunity defense. The court reasoned that a website "can be [simultaneously] both a service provider and a content provider"[56] and that "[b]y requiring subscribers to provide the information as a condition of accessing its service, and by providing a limited set of prepopulated answers, Roommate bec[ame] much more than a passive transmitter of information provided by others; it bec[ame] the developer, at least in part, of that information."[57] In so holding, the court departed slightly from *Zeran*'s extremely narrow articulation of what constitutes "content development." Instead, the court recognized that the range of potential liability must extend somewhat further than literal authorship to other parties who are also directly responsible. Roommates.com was responsible because it solicited and aided the content's development. That the user posting to the site was the last in the chain to push a button or click a mouse was insignificant because the content was the product of a collaborative effort.[58]

Cases such as *Roommates.com* are rare but not unheard of.[59] They represent a push against the broad immunity of the majority view and make some effort to address the interpretive difficulties that courts have faced in applying Section 230.[60] What these courts have failed to do, however, is to provide a theoretical foundation that explains their deviations from the majority view and that defines the precise boundaries of the exceptions they create to the general rule of immunity for providers of third-party-created content. They announce "no immunity here" without providing a satisfactory account of how Section 230 should now be understood.

---

55. *Roommates.com*, 521 F.3d at 1162. The Fair Housing Act prohibits discrimination on the basis of "race, color, religion, sex, familial status, or national origin," 42 U.S.C. § 3604(c) (2006), and the California fair housing law prohibits discrimination on the basis of "race, color, religion, sex, sexual orientation, marital status, national origin, ancestry, familial status, source of income, or disability." CAL GOV'T CODE § 12955 (West 2005).

56. *Roommates.com*, 521 F.3d at 1162.

57. *Id.* at 1166.

58. *Id.* at 1166–67.

59. *See, e.g.*, MCW, Inc. v. Badbusinessbureau.com, L.L.C., No. Civ. A.3:02-CV-2727-G, 2004 WL 833595 (N.D. Tex. Apr. 19, 2004); *see also* Chi. Lawyers' Comm. for Civil Rights Under Law, Inc. v. Craigslist, Inc., 519 F.3d 666, 670–71 (7th Cir. 2008) (rejecting majority view of Section 230 immunity and instead interpreting Section 230 immunity to apply only to publication-based torts); Batzel v. Smith, 333 F.3d 1018, 1033 (9th Cir. 2003) (reasoning that a defendant would not be immune under Section 230 where, though the content was provided by a third party, it was not provided with the expectation that it would be posted to the Internet).

60. *See supra* Part III.

This Part attempts to provide that missing framework by analyzing Section 230 in light of the relationship between two strains of pre-Internet vicarious liability defamation doctrine and *Stratton Oakmont, Inc. v. Prodigy Services Co.*,[61] the defamation case that prompted Congress to enact Section 230. The analysis indicates that although the immunity provision of Section 230 is broad, Congress did not intend to abrogate entirely traditional common law notions of vicarious liability. Some bases of vicarious liability remain, and their continuing validity produces a cogent distinction between "content providers" and "content developers."[62] It also resolves the tension between Section 230's encouragement of "Good Samaritan" screening and its simultaneous provision for sweeping immunity regardless of any censorial action.[63]

### A. *Pre-Internet Precursors: Ratification and Concert of Action*

The common law of defamation provides two relevant lenses through which to interpret Section 230 immunity. First, consider the pre-Internet theory of defamation by ratification. Under that doctrine, "[o]ne who intentionally and unreasonably fails to remove defamatory matter that he knows to be exhibited on land or chattels in his possession or under his control is subject to liability for its continued publication."[64] These cases typically involve a defendant who, though not the author of the defamatory statement in question, has implicitly ratified that statement by his failure to remove it from a place of prominence on his property. In *Tacket v. General Motors Corp.*,[65] for example, the Court of Appeals for the Seventh Circuit considered the defamation claim of Thomas Tacket, a General Motors employee, against that company for its eight-month failure to remove an unauthorized and allegedly defamatory stenciling from its factory walls. Tacket had been im-plicated in a scheme in which he, in his managerial capacity with General Motors, had approved a contract to purchase wooden boxes from a firm located in his friend's garage.[66] Upon learning of the scheme, Tacket's fellow employees expressed their displeasure by inscribing the phrase "TACKET TACKET WHAT A RACKET" on a highly visible area of a factory wall.[67] The Seventh Circuit held that the message could not have remained on the wall for eight months without approval of the company's management and that the company had therefore adopted the statements as its own.[68] Though somewhat rare, defamation by ratification is a natural extension of basic defamation principles. By adopting statements of a third party as his own and continuing to publish them on his property, a defendant has taken individual ownership of these statements. They belong as much to him as if directly spoken, and they are no less damaging.

---

61. No. 31063/94, 1995 WL 323710 (N.Y. Sup. Ct. May 24, 1995).
62. *See supra* Part III.B.
63. *See supra* Part III.A.
64. *See* RESTATEMENT (SECOND) OF TORTS § 577(2) (1977).
65. 836 F.2d 1042 (7th Cir. 1987).

Second, consider defamation by concert of action.[69] As in criminal law, defendants who cooperate to pursue a tortious goal may be held liable together, even though each has not individually committed every element of the offense. If two parties target a plaintiff's house for robbery, and one breaks down the door while the other beats the plaintiff, each is liable for his own acts as well as those of his coconspirator.[70] Similarly, in the defamation context, if several actors join together to create and disseminate defamatory material, all may be held liable. For example, in *Gosden v. Louis*,[71] the Court of Appeals of Ohio considered the defamation claim of a construction company against seventeen residents of an apartment complex where the company had performed work.[72] One of the residents had drafted a letter, later signed by the other residents, alleging that while working at the complex, employees of the company had driven their vehicles recklessly, harassed neighborhood residents, used profane language, engaged in lewd and voyeuristic acts, and violated a county noise ordinance.[73] Each of the defendants freely admitted to having signed the letter, and the court held these admissions sufficient to support a jury finding of conspiracy to defame the plaintiff.[74] Like ratification torts, concert of action torts are relatively rare. Nonetheless, the category is a logical expansion of the tort doctrine that holds all participants in tortious schemes equally liable. Tortfeasors should not escape liability simply because they choose to work in groups.

---

66. *Id.* at 1044.

67. *Id.* at 1045, 1047.

68. *Id.* at 1047.

69. For a general treatment of contributing tortfeasor (or "civil conspiracy") liability, see RESTATEMENT (SECOND) OF TORTS § 876: "For harm resulting to a third person from the tortious conduct of another, one is subject to liability if he (a) does a tortious act in concert with the other or pursuant to a common design with him, or (b) knows that the other's conduct constitutes a breach of duty and gives substantial assistance or encouragement to the other so to conduct himself . . . ." *Id.*

70. *See* DOBBS, *supra* note 8, § 340; *see also* Drake v. Keeling, 299 N.W. 919 (Iowa 1941); Smithson v. Garth, 3 Lev. 323, 83 Eng. Rep. 711 (1691).

71. 687 N.E.2d 481 (Ohio Ct. App. 1996).

72. *Id.* at 486–87.

### B. A Narrower Interpretation of Section 230

The tort theories of ratification and concert of action help to explain the perplexing relationship between subsections (c)(1) and (c)(2) of Section 230.[75] Subsection (c)(1) provides that "[n]o provider or user of an interactive computer service shall be treated as the publisher or speaker of any information provided by another information content provider."[76] Because this language seems to provide complete immunity to service providers regardless of whether they choose to censor content or not, the subsection appears inconsistent with (c)(2), which provides immunity only to service providers that do choose to censor content.[77]

Ratification and concert of action relieve the tension between the two subsections. Despite its strong language, subsection (c)(1) cannot have been intended to immunize service providers from liability for third-party-created content in every context. Such an interpretation would yield nonsensical results. Congress could not have intended, for instance, to immunize a website run by a political organization that conspires with a third-party author to defame an opposing candidate viciously. The strict application of subsection (c)(1)'s language appears to confer such immunity, but that language must be interpreted in light of the background principles of tort law.[78] Something remains of

---

73. *Id.*
74. *Id.* at 498.
75. *See supra* Part III.A.
76. 47 U.S.C. § 230(c)(1) (2006).
77. *See supra* Part III.A.
78. Apart from the absurd results that follow an interpretation providing immunity in absolutely every instance of third-party-created content, an interpretation that preserves traditional common law bases of vicarious liability can be justi-liability for third-party-created content even after subsection (c)(1): liability where a website either ratifies content created by a third-party or is a coconspirator in its creation. Once it is understood that subsection (c)(1)'s immunity is not absolute—that its language does not abrogate traditional bases of vicarious liability[79]—subsection (c)(2)'s purpose becomes much clearer.

The purpose of subsection (c)(2) is to immunize websites from certain types of vicarious liability. Section 230 was drafted partly in response to *Stratton Oakmont, Inc. v. Prodigy Services Co.*,[80] which found an ISP liable because, despite a stated policy of censoring objectionable content, the company failed to censor statements that falsely accused Stratton Oakmont and its president of crimi-

---

fied in another way as well: Ratification liability and, to a lesser extent, concert of action liability are based on the idea that tortious action belongs not only to the immediate tortfeasor but also to the vicariously liable actor. This is easy to see in ratification liability. Indeed, the word "ratification" suggests the principle. By approving or sanctioning the actions of another, one takes ownership of those actions such that it is no different than if one had committed them personally. Similarly, conspiracy liability is based on the idea that liability flows from the tortious course of conduct taken as a whole, and that each individual participant is responsible for the entirety of the conduct as if each aspect were individually undertaken. Viewed from this perspective, ratified content or coconspirator-created content is not really third-party-created content at all. It is content created by the service provider vicariously.

79. Indeed, no one could reasonably argue that Section 230 was intended to eliminate *all* bases of vicarious liability. There is no doubt, for instance, that Section 230 left intact the doctrines of respondeat superior and agency liability. A website owner who argued that he should receive immunity because it was his employee, a third party, who posted tortious material would be laughed out of court. All courts to consider the issue have simply assumed that standard principles of agency law remain unchanged by Section 230's immunity provision. *See* Batzel v. Smith, 333 F.3d 1018, 1035 (9th Cir. 2003) ("Agency is the fiduciary relationship that arises when one person (a 'principal') manifests assent to another person (an 'agent') that the agent shall act on the principal's behalf and subject to the principal's control, and the agent manifests assent or otherwise consents so to act. In order for [the defendant] to be held vicariously liable for the torts of [the third party tortfeasor] on a theory of agency, [the defendant] must have had the ability to control [the third party's] activities." (internal citations omitted)); Higher Balance, LLC v. Quantam Future Group, Inc., No.08-223-HA, 2008 WL 5281487, at *7 (D.Or. 2008) ("[P]laintiff has failed to show that the [website] forum moderators are . . . staff members. Without this evidentiary link, plaintiffs have not shown that the forum moderators are employees or agents . . . ."); Blumenthal v. Drudge, 992 F. Supp. 44, 50 (D.D.C. 1998) ("It is also apparent to the Court that there is no evidence to support the view originally taken by plaintiffs that [the third party tortfeasor] is or was an employee or agent of [the defendant]."). Ratification and concert of action are no different.

80. No. 31063/94, 1995 WL 323710 (N.Y. Sup. Ct. May 24, 1995); *see also supra* Part I.

nal activity.[81] The theory of liability in *Stratton* was one of ratification. The court reasoned that "Prodigy ha[d] uniquely arrogated to itself the role of determining what [was] proper for its members to post and read"[82] and could therefore be held liable as a publisher of any defamatory material posted to its bulletin board system. By failing to remove defamatory statements from its system, Prodigy had ratified and republished those statements, making it no less liable than the original speaker.[83] With subsection (c)(2), Congress essentially overruled *Stratton*, stipulating that failure to censor never constitutes ratification. Congress recognized that service providers would be discouraged from censoring at all if by doing so they risked incurring liability and so stepped in and disqualified censorship as a ground for ratification liability.

Other avenues for vicarious liability nevertheless remain even after subsection (c)(2). Websites and other service providers can still be liable for third-party-created content if they ratify that content in a manner other than by failure to censor, and they can still be liable if they engage in a tortious concert of action with a third party. As an example, consider the website of the Republican party, gop.com. If an Internet user posts content calumniously accusing President Obama of treason or of piracy on the high seas, gop.com will be immune from defamation liability under subsection (c)(2). Suppose, though, that instead of merely failing to remove the post, gop.com reprogrammed its website such that the post was prominently displayed to every visitor of the site. Alternatively, suppose that the post was part of a larger right-wing conspiracy of which gop.com was a part—that gop.com had acted in concert with the defamer to make his voice heard. Under such circumstances, gop.com's invocation of Section 230 immunity should fail. Though not the creator of the content, gop.com could still be liable as a ratifier or coconspirator of the defamer.

Principles of vicarious liability also clarify another difficulty courts have faced in applying Section 230: the service provider-content provider distinction.[84] What does it mean to be "responsible, in whole or in part, for the creation or development"[85] of

---

81. *Stratton*, 1995 WL 323710, at *4.
82. *Id.*
83. *See id.*; RESTATEMENT (SECOND) OF TORTS §§ 577(2), 578 (1977)(including failure to remove defamatory materials in the definition of publication).
84. *See supra* Part III.B.
85. 47 U.S.C. § 230(f)(3) (2006).

Internet content, and when do websites cross the line and become content providers? The slight trend toward broader liability is a push in the right direction, but courts have thus far insufficiently articulated the boundaries of expanded liability. The pre-Internet theories of ratification and concert of action liability help to define those boundaries. Responsibility should extend somewhat further than the *Zeran* line seems to allow. The category of individuals responsible for the creation or development of content is broader than literal content authors. It includes not only the last individual in a chain to press a button or click a mouse[86] but also all of the ratifiers[87] and coconspirators along the way. A few courts have, admirably, moved to narrow Section 230 immunity. But, without a coherent interpretive framework by which to distinguish who is and who is not "responsible in part" for the "creation or development" of content, they leave themselves open to the critique that they have merely carved out exceptions by judicial fiat—that there is no principled way to deny immunity in one case but not another.[88] The common law doctrines of conspiracy and ratification can supply the basis for those distinctions. Their application will allow courts to make principled determinations as to what sorts of conduct take service providers out of the realm of immunity and into the realm of responsibility for objectionable content.

CONCLUSION

This historical, common law-focused approach to the scope of Section 230 immunity accomplishes several goals. First, by recognizing that two categories of vicarious liability survive subsection (c)(1), it is possible to explain the nagging question of subsection (c)(2)'s purpose given subsection (c)(1)'s seemingly endless breadth. Subsection (c)(2) explicitly eliminates the rationale for ratification liability relied on in *Stratton v. Oak-*

---

86. *See* Fair Hous. Council of San Fernando Valley v. Roommates.com, LLC, 521 F.3d 1157, 1166–67 (9th Cir. 2008) ("The projectionist in the theater may push the last button before a film is displayed on the screen, but surely this doesn't make him the sole producer of the movie.").

87. Excluding, of course, ratification by failure to censor. Such theories of ratification are expressly prohibited by subsection (c)(2).

88. *See, e.g.*, *Roommates.com*, 521 F.3d at 1183 (McKeown, J., concurring in part and dissenting in part) ("The majority's definition of 'development' would transform every interactive site into an information content provider and the result would render illusory any immunity under § 230(c).").

*mont, Inc. v. Prodigy Services Co.*—ratification by failure to censor. Second, by articulating the ways in which a party who is not the literal author of content may nonetheless be responsible for that content, it is possible to clarify slightly the sometimes difficult distinction between content providers and mere service providers. Finally, by grounding the content provider-service provider distinction in the common law doctrines of ratification and concert of action, it is possible to provide a theoretical underpinning to the nascent trend toward narrower Section 230 immunity. Courts are empowered to apply a more developed framework to questions of responsibility, avoiding accusations that the concept is limitless or a creation of judicial fiat.

*Gregory M. Dickinson*